\documentclass[aps,prl,twocolumn]{revtex4}
\usepackage{graphicx}
\usepackage{latexsym}
\usepackage{amsmath}
\usepackage{amsfonts}
\usepackage{amssymb}
\usepackage{color}
\usepackage{units}
\usepackage{natbib}
\newcommand{\be}{\begin{equation}}
\newcommand{\ee}{\end{equation}}
\newcommand{\bea}{\begin{eqnarray}}
\newcommand{\eea}{\end{eqnarray}}
\usepackage[parfill]{parskip}
\usepackage{orcidlink}
\begin{document}
\title{Correlated and anti-correlated density dependent motility} 
\author{Itay Azizi \orcidlink{0000-0003-2939-4421}} 
\address{Correspondence: itay.azizi@gmail.com}
\begin{abstract} 
I study via Langevin dynamics simulations two opposite cases of systems of particles that alternate their identity according to density dependent motility (DDM) rules and interact via a soft repulsive potential. In the correlated case, dilute regions are passive and dense regions are active, while in the anti-correlated case, dilute regions are active and dense regions are passive. I classify the emerging steady states, explain the principal phase transitions, and finally suggest directions for further investigation.
\end{abstract} 
\maketitle 
\section{Introduction}
\begin{figure}[ht] 
\includegraphics[width=0.95\linewidth]{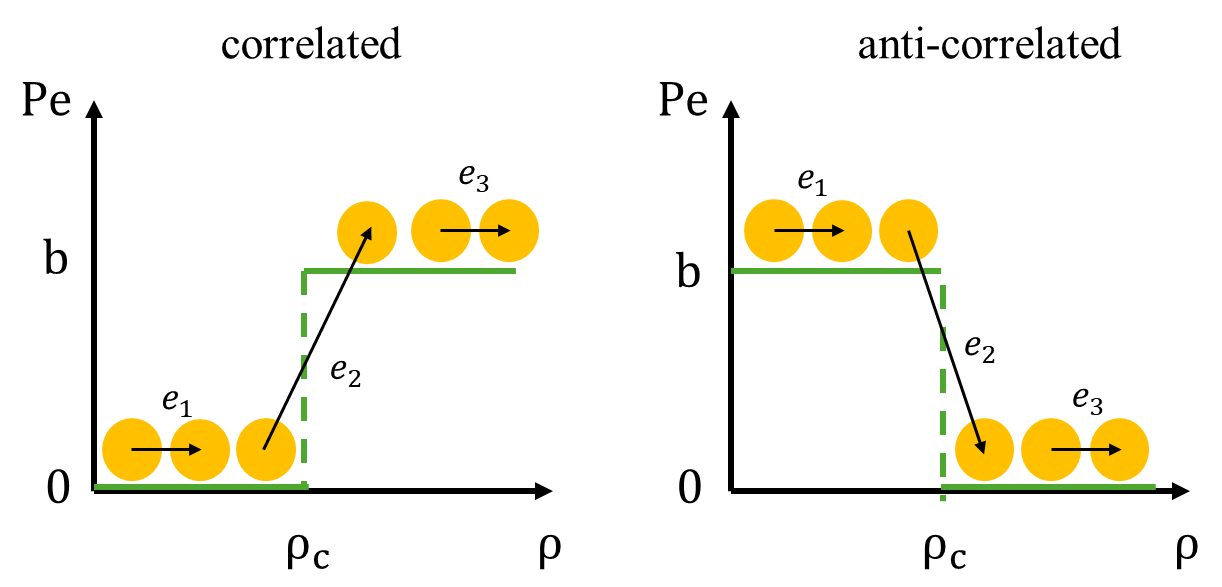}
	\caption{\label{fig:types} Correlated versus anti-correlated density dependent motility. In the correlated case, dilute regions are passive and dense regions are active. In the anti-correlated case, dilute regions are active and dense regions are passive. When the local density of a particle increases, it can participate in one of three events, marked by $e_1$, $e_2$ and $e_3$.}
\end{figure}

Quorum sensing is an important biological phenomenon that describes the ability of a single bacterial cell to sense the number of cells around it (cell density) such that a group of cells can react to changes in density with a coordinated response. 
Quorum sensing dominates different behaviors that are relevant to physiological functions, such as swarming motility, biofilm maturation and antibiotic resistance \cite{Whiteley2017}. Recently, two component systems of particles that change their motility with local density were used as a toy model for quorum sensing \cite{Bauerle2018,Fischer2020,Souza2025}. 
These works demonstrate that hard spheres with discontinuous DDM form diverse phases as a function of activity, critical density and quorum sensing range. For example, clusters with different shapes, vortex clusters, active gels and long-lived transients. Phase separation can occur at lower packing fraction than the typical values for a purely active system \cite{Jose2021}. 
\\As a generic argument, in a quorum sensing model, motility can be correlated or anti-correlated with local density as described in Fig.\ref{fig:types}. 
In both cases the distinction between dilute and dense is in respect to critical density $\rho_c$ and the change in motility is discontinuous. 
In the correlated case, particles are passive in dilute regions and active at dense regions, and in the anti-correlated case, particles are active at dilute regions and passive at dense regions. 
While the aforementioned works focused on systems of particles that interact via a hard core potential in the anti-correlated case \cite{Bauerle2018,Fischer2020,Jose2021,Souza2025}, they did neither explore the correlated case, nor other interaction potentials.   
\\The two cases describe different systems. 
The anti-correlated case matches to organisms that their motion is suppressed in dense regions, such as social insects. 
In the case of bees, the structure and morphology of their clusters were studied decades ago experimentally \cite{bees1954} and later theoretically \cite{bees2000}. 
In the case of ants, it was shown that quorum requirement can help a colony of ants choose the best available site when they emigrate due to damaged nest \cite{ants1} and there is recruitment by tandem runs \cite{ants2}. 
On the contrary, the correlated case matches to organisms that their motion is enhanced at dense regions, such as Dictyostelium whose single 'diluted' cells walk randomly and when density reaches a threshold, their motion becomes highly coordinated and that enables phototaxis \cite{Nizak2025}. 
Further relevance is outside the scope of biology, for social systems of humans or robots with decentralized decision-making. 
Since biological cells and organisms are deformable, a soft potential can represent the interactions between them better than a hard potential. 
\\Therefore, I utilize molecular dynamics simulations to answer the novel question: what are the emerging phases in systems of particles with correlated/anti-correlated DDM that interact via a soft repulsive potential?  
\section{Methods}
In order to elucidate the behavior of systems with density dependent motility, I carried out Langevin dynamics simulations using a code I wrote in Fortran. 
My two-dimensional system consists of $N=2025$ particles in a square box with periodic boundary conditions and dimensions $L=L_x=L_y=82.16$ which sets a global density of $\rho=0.3$. $k_B=1$ and $T=1$. 
Particles interact via a short range repulsive potential 
\begin{equation}
V(r) = \epsilon r^{-6}
\end{equation} 
$\epsilon=36.5$ such that at the cutoff distance $r_c=3$, the amplitude of interaction forces is negligble compared to the amplitude of thermal fluctuations. 
This way too weak repulsive interactions are not calculated unnecessarily. 
The choice in a purely repulsive potential is to associate any clustering behavior with quorum sensing rather than direct attractive interactions. 
\\ Particle $i$ can alternate between active and passive identity as a function of its local density $\rho_i$ that is defined in reference to the natural lengthscale of the system, namely the mean interparticle distance $a=\rho^{-1/2}$. 
The code counts $m_i$: the number of neighbors of particle $i$ within a disk of radius of $2a$ whose center is particle $i$. 
The area of this disk is $A$ and the local density is $\rho_i=m_i/A$. $2a$ indicates the range of action of the quorum sensing.
\\ Particle $i$ is assigned with a position $r_i$ and an orientation $\theta_i$ that evolve according to Langevin equations: 
\begin{eqnarray}
\frac{d\mathbf{r}_i(t)}{dt} &=& v_p(\rho_i) \mathbf{e_i}(t) - \beta D \nabla U_i(t) + \sqrt{2D} \, \boldsymbol{\eta}(t), \label{eom1} \\
\frac{d\theta_i(t)}{dt} &=& \sqrt{2D_r(\rho_i)} \, \xi(t)
\label{eom2}
\end{eqnarray}
where $v_p$ is the magnitude of the self-propulsion velocity in the direction of the particle $\mathbf{e}_i(t)=(\cos\theta_i(t), \sin\theta_i(t))$. 
The Péclet number (\text{Pe}) is defined as the ratio of persistence length $l_p=v_p \tau_r$ : how far an active particle moves before its orientation randomizes due to rotational diffusion with $\tau_r=1/D_r$, to diffusive length $l_D=\sqrt{D \tau_r}$: how far this particle diffuses during this time. 
\begin{equation}
\text{Pe} = \dfrac{l_p}{l_D} = \dfrac{v_p}{\sqrt{DD_r}}.
\label{pe}
\end{equation}
The translational (rotational) diffusion coefficient is denoted by $D$ ($D_r$), and both $\boldsymbol{\eta}(t)$ and $\xi(t)$ are Gaussian white noise terms satisfying   
$\langle \eta_i (t) \eta_j (t^{\prime}) \rangle = \delta_{ij} \delta(t-t^{\prime})$ with $i,j \in (x,y)$ and  $\langle \xi(t) \xi(t^{\prime}) \rangle = \delta(t-t^{\prime})$, respectively. $\gamma$ and $\gamma_r$ are the translational and rotational friction coefficients, satisfying $D=k_BT/\gamma$ 
and $D_r=k_BT/\gamma_r$, respectively. $\gamma=10$ and $\gamma_r=4.2$.
\\$U$ denotes the potential energy arising from conservative forces, e.g., pair interactions such that $U_i = \sum_{j \neq i} V(\vec r_j - \vec r_i)$, and the gradient is with respect to the position of particle $i$. 
\\According to local density, for a passive particle, $v_p=Pe=D_r=0$ and for an active particle $v_p$ is chosen such that $Pe=b$. See in Fig.\ref{fig:types}, the possible events for an individual particle as local density increases. 
In the correlated case, $e_1: passive \rightarrow passive$,  $e_2: passive \rightarrow active $ and $e_3: active\rightarrow active$. 
In the anti-correlated case, $e_1: active \rightarrow active$,  $e_2: active \rightarrow passive $ and $e_3: passive \rightarrow passive $. 
\\Therefore, each simulation studies one of the two cases at a single point of phase space: a specific critical density $\rho_c$ in the range $[0.15,0.45]$ and a specific activity $b$ in the range $[5,50]$. 
The simulation begin from a square lattice and run up to $10^6$ steps with a timestep of $\Delta t=0.005$. 
This runtime is sufficient to reach a steady state that is characterized by time independent parameters, such as potential energy, active particles fraction, and clusters size and morphology.

\section{Results}
\subsection{Part 1: The Correlated Case}
\subsubsection{Visual Inspection}
\begin{figure}[ht] 
\includegraphics[width=0.95\linewidth]{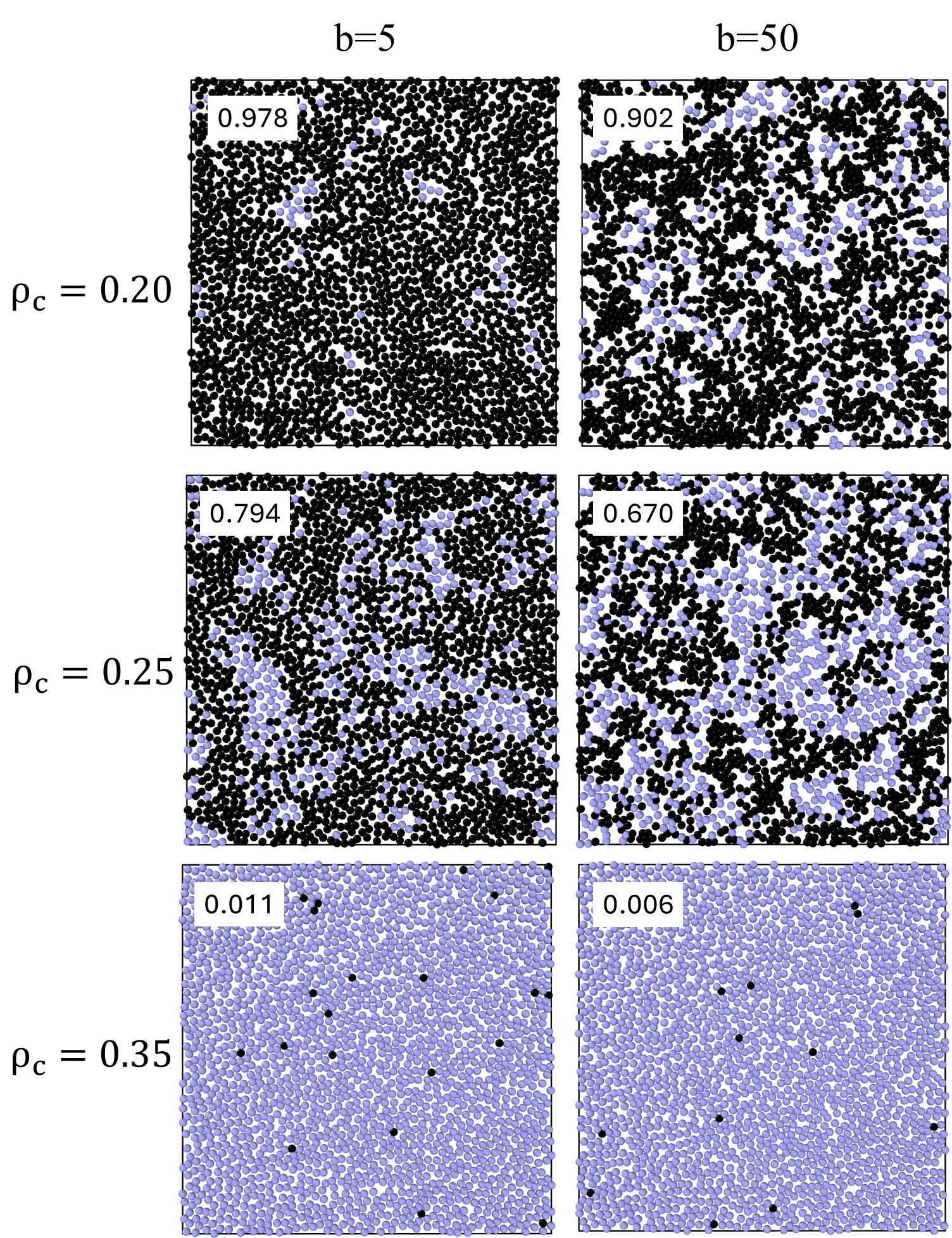}
\caption{\label{fig:CORR_snapshots} Snapshots for the correlated case at $b=5, 50$ and $\rho_c=0.20, 0.25, 0.35$. Passive (active) particles are colored in gray (black). The fraction of active particles is indicated on the top-left corner of each snapshot.}
\end{figure}

According to the model, at very low (high) critical density, there is obviously an active (passive) fluid. 
At intermediate critical density, I observe a fluid-fluid phase separation: dense clusters of active particles that coexist with a dilute passive fluid. 
The active clusters are metastable and change their members and shape with time. 
See Fig.\ref{fig:CORR_snapshots} for the three states described above. 
\\Therefore, the two principal transitions are: $t_1:$ "active fluid" $\rightarrow$ "fluid-fluid" and $t_2:$ "fluid-fluid" $\rightarrow$ "passive fluid". 
In the "fluid-fluid" state, the interesting events are $e_2$ and $e_3$ that correspond to a passive particle entering an active cluster, and motion of active particle inside an active cluster, respectively. 
\subsubsection{Quantitative Analysis}
\begin{figure}[ht]  
\includegraphics[width=0.95\linewidth]{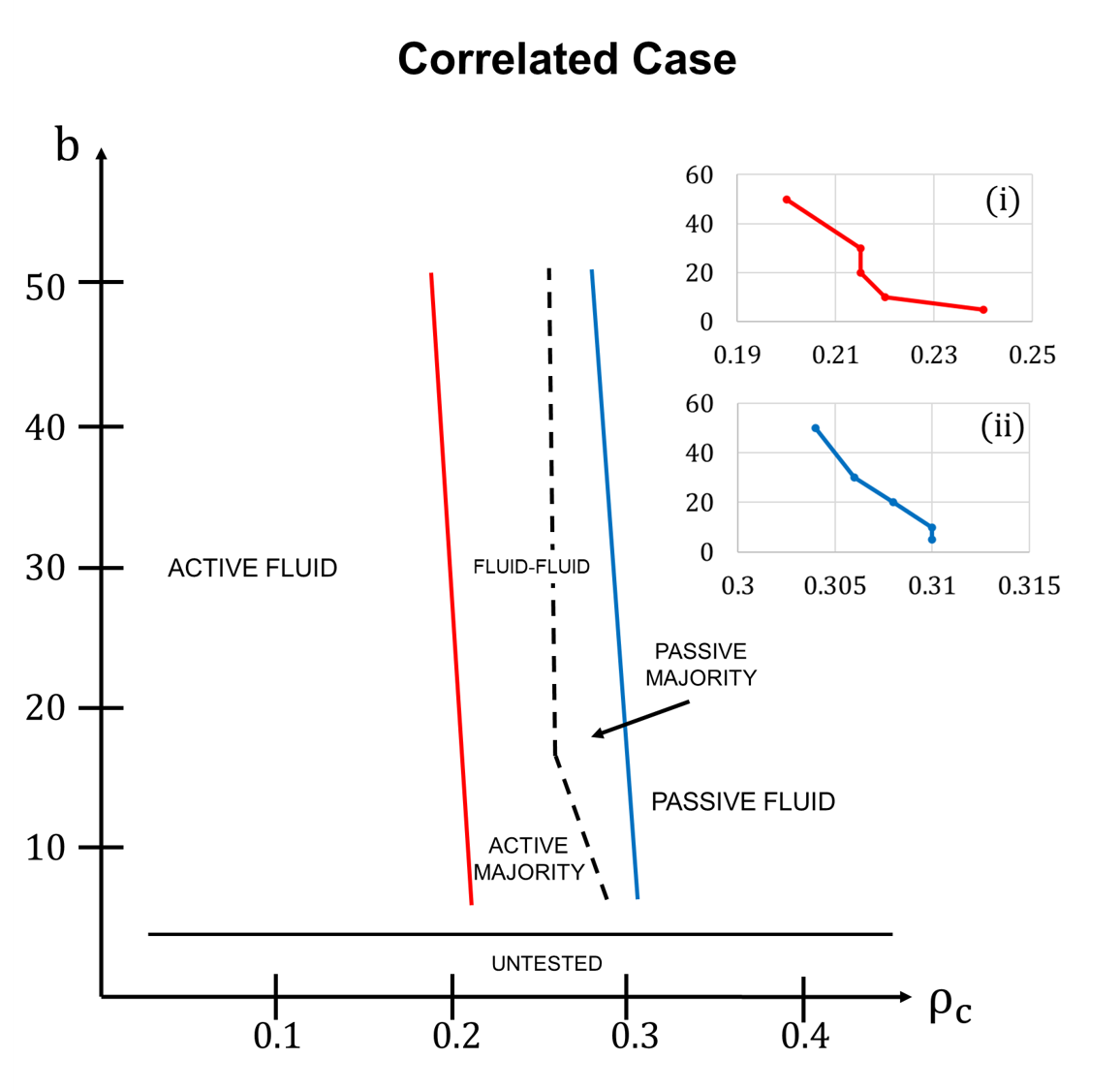}
	\caption{\label{fig:CORR_DIAGRAM} Phase diagram for the correlated case. The transition from "active fluid" to "fluid-fluid" is marked by a red line and the transition from "fluid-fluid" to "passive fluid" is marked by a blue line. The sub-transition from "active majority" to "passive majority" is marked by the black dashed line. Insets (i) and (ii) show the tendency of the principal phase lines.}
\end{figure}

The location of each transition can be estimated via a simple measurement of the active fraction, $\phi_a=<N_a>/N$, where $<N_a>$ is the average number of active particles in steady state. 
See Fig.S1 in Supplemental Material (SM) for the active fraction as a function of critical density. 
$t_1$ coincides with a small reduction in active population, $\phi_a=0.9$, and $t_2$ coincides with a greater decrease, $\phi_a=0.1$. In Fig.\ref{fig:CORR_DIAGRAM} the red line matches to $t_1$ and the blue line to $t_2$. 
See in the insets, the magnified phase lines which uncover an interesting effect: as $b$ increases, the first transition as well as the second transition happen at lower $\rho_c$. 
This indicates that in the first transition activity promotes phase separation, similar to motility induced phase separation (MIPS), but in the second transition activity leads the particles to more dilute regions and they become passive. 
Hence, high activity produces two effects, in one regime it supports phase separation and te second regime it supports a pure phase phase.  
\\The structure of the active phase is studied using the radial distribution function of the active particles, $g_{aa}(r)$. For example, at $b=30$, $g_{aa}(r)$ is plotted for different values of $\rho_c$ in Fig.S2 in the SM. As $\rho_c$ increases, the first peak increases which means that the effective attractive attractions between the active particles are stronger.
\begin{figure}[ht]  
\includegraphics[width=1.0\linewidth]{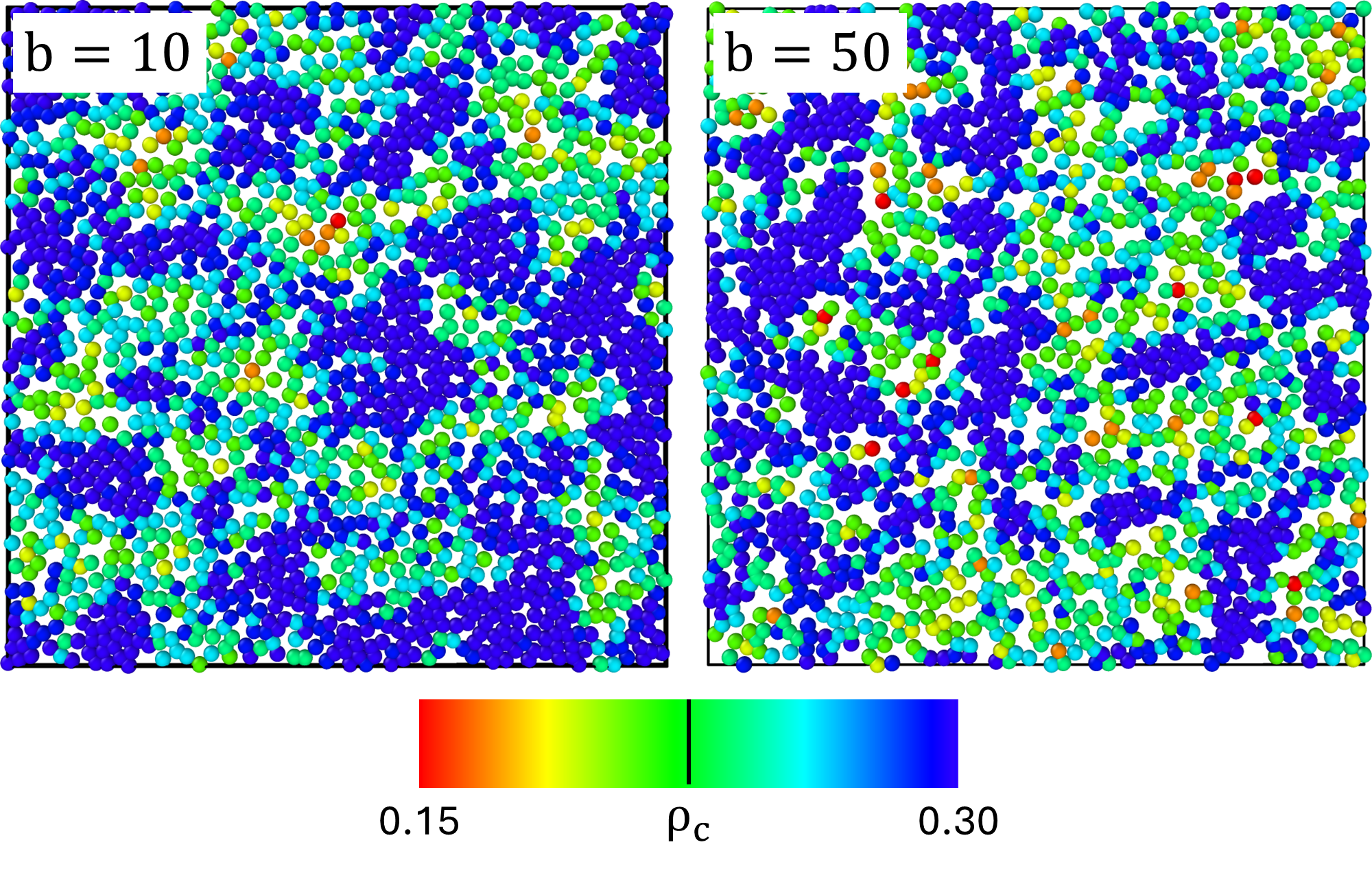}
	\caption{\label{fig:CORR_COLOR} Correlated case: systems in the "active majority" state at $\rho_c=0.225$. Particles colored according to local density as indicated in the colormap. As $b$ increases, the active clusters are smaller.}
\end{figure}
\\Another way of showing the how activity affects active clusters is by coloring the particles according to local density in the vicinity of $\rho_c$, see Fig.\ref{fig:CORR_COLOR}. 
As activity increases, the size of the active clusters decreases. 
\\ Within the "fluid-fluid" phase, I indicated an additional transition when the active fraction crosses $0.5$, the system has a sub-transition: 
"active majority" $\rightarrow$ "passive majority" that is marked in the phase diagram with a dashed black line.  
\\ At the global density of $\rho=0.3$, I did not observe hexatic active fluid. 
I was running additional simulations at a higher global density of $\rho=0.5$ and also there did not observe an hexatic active phase (not shown). 
\begin{figure}[ht] 
\includegraphics[width=0.95\linewidth]{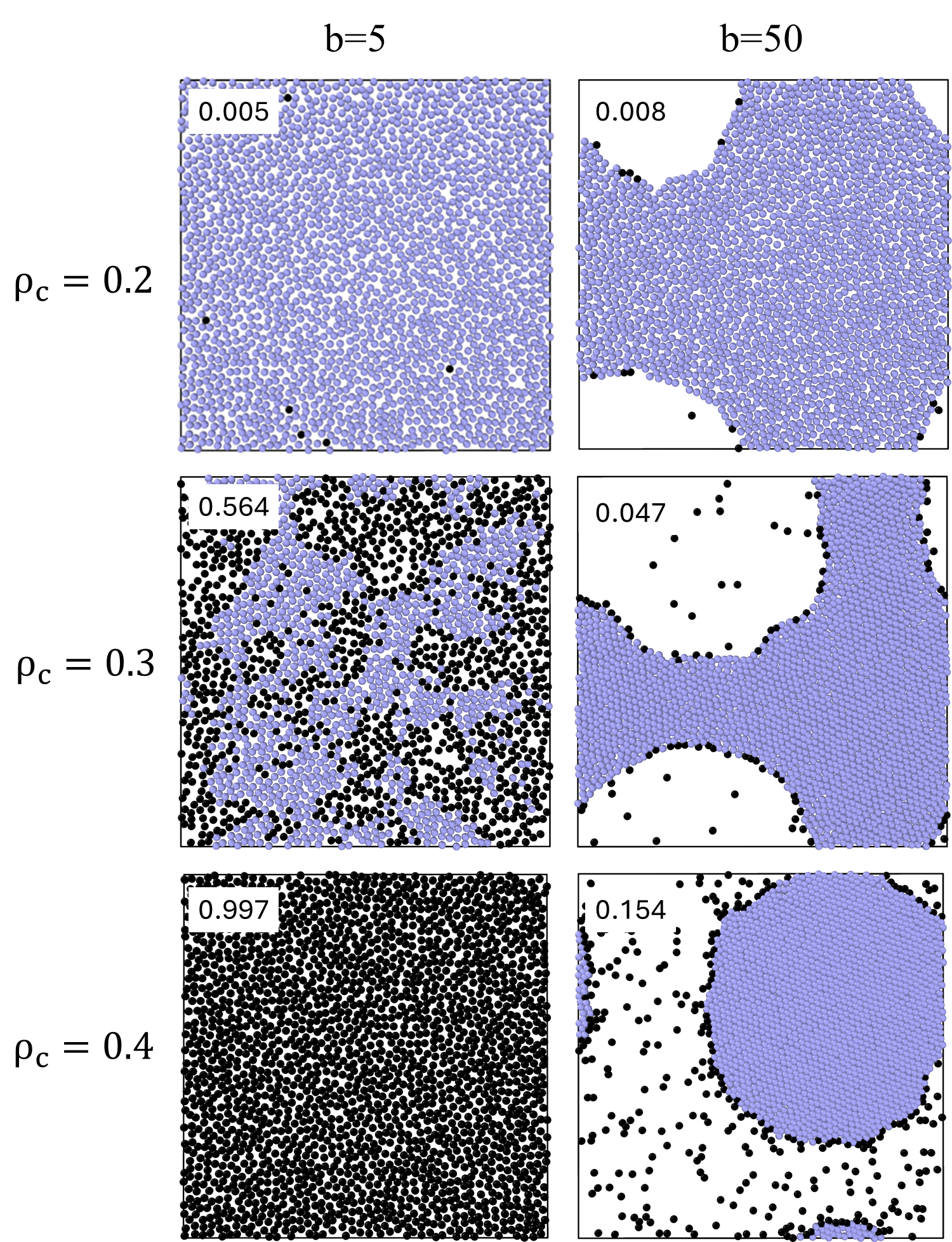}
	\caption{\label{fig:AC_snapshots} Snapshots for the anti-correlated case $b=5, 50$ and $\rho_c=0.2, 0.3, 0.4$. Passive (active) particles are colored in gray (black). The fraction of active particles is indicated on the top-left corner of each snapshot. The system gradually phase separates into a cluster of passive particles coexisting with an active gas.}
\end{figure}
\subsection{Part 2: The Anti-Correlated Case}
\subsubsection{Visual Inspection}
According to the model, at very low (high) local density, there is obviously a passive (active) fluid. 
Therefore, the intriguing behavior should appear at intermetdiate local density near the global density of the system ($\rho=0.3$). 
\begin{figure}[ht] 
\includegraphics[width=0.95\linewidth]{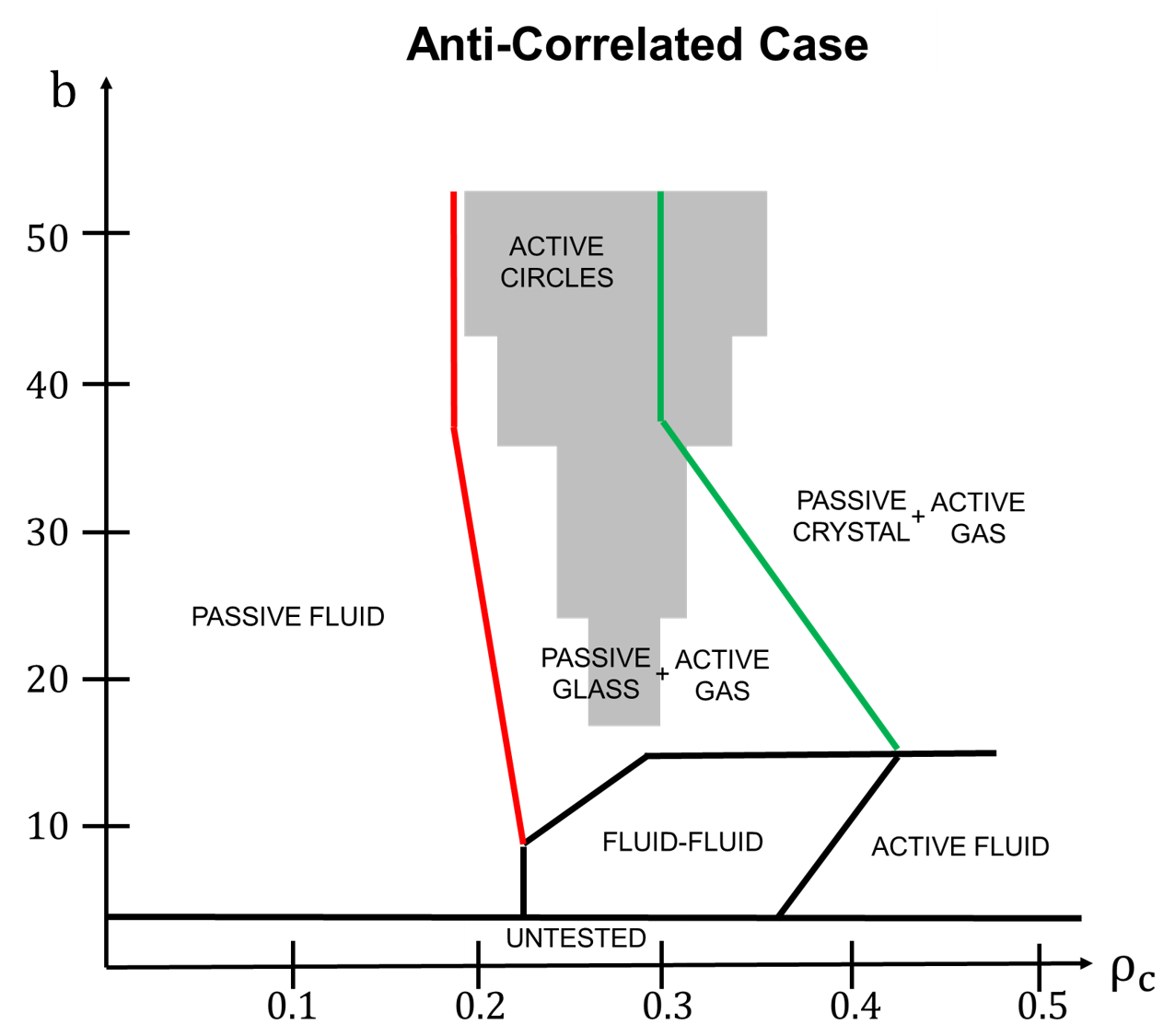}
	\caption{\label{fig:AC_DIAGRAM} Phase diagram for the anti-correlated case. The transition from ”passive fluid” to ”glass-fluid” is marked by a red and the transition from ”glass-fluid” to ”crystal-fluid” is marked by a green line. The sub-state of "active circles" is marked by the gray zone.}
\end{figure}
In Fig.\ref{fig:AC_snapshots}, I show snapshots that indicate different states at intermediate local density. 
While at a low activity, e.g., $b=5$, there is segregation into a fluid-fluid state, e.g., $b=50$, there is a phase separation into a fluid-solid. 
Interestingly, the passive solid can be amorphous or hexatic and the active fluid can be dispersed or form circular clusters of swimmers. 
The interface between a passive solid to active fluid can contain both active and passive particles. 
For example, see Fig.\ref{fig:AC_snapshots} at $b=50$, $\rho_c=0.3$ that the swimmers press the passive particles into a compact hexatic phase. 
The solidification of the passive particles keeps them deeper in the passive regime. 
However, at a lower value of $\rho_c$, for example at $b=50$, $\rho_c=0.2$, this pressure is lower and the passive solid is amorphous (glass).
\\For activity $b$ in the range $[10,50]$, as $\rho_c$ increases, there are three principal transitions: $t_1:$ "passive fluid" $\rightarrow$ "passive glass+active gas",  $t_2:$ "passive glass+active gas" $\rightarrow$ "passive crystal+active gas", and $t_3:$ "passive crystal+active gas" $\rightarrow$ "active fluid". 
In the state "passive glass+active gas", the interesting events are $e_2$ and $e_3$ that correspond to an active particle entering the glassy phase and motion inside the glassy phase, respectively. 
In the state "passive crystal+active gas", $e_2$ can correpond also to surface particles that change their identity from active to passive, and vice versa. 
\subsubsection{Quantitative Analysis}
\begin{figure}[ht]
\includegraphics[width=0.6\linewidth]{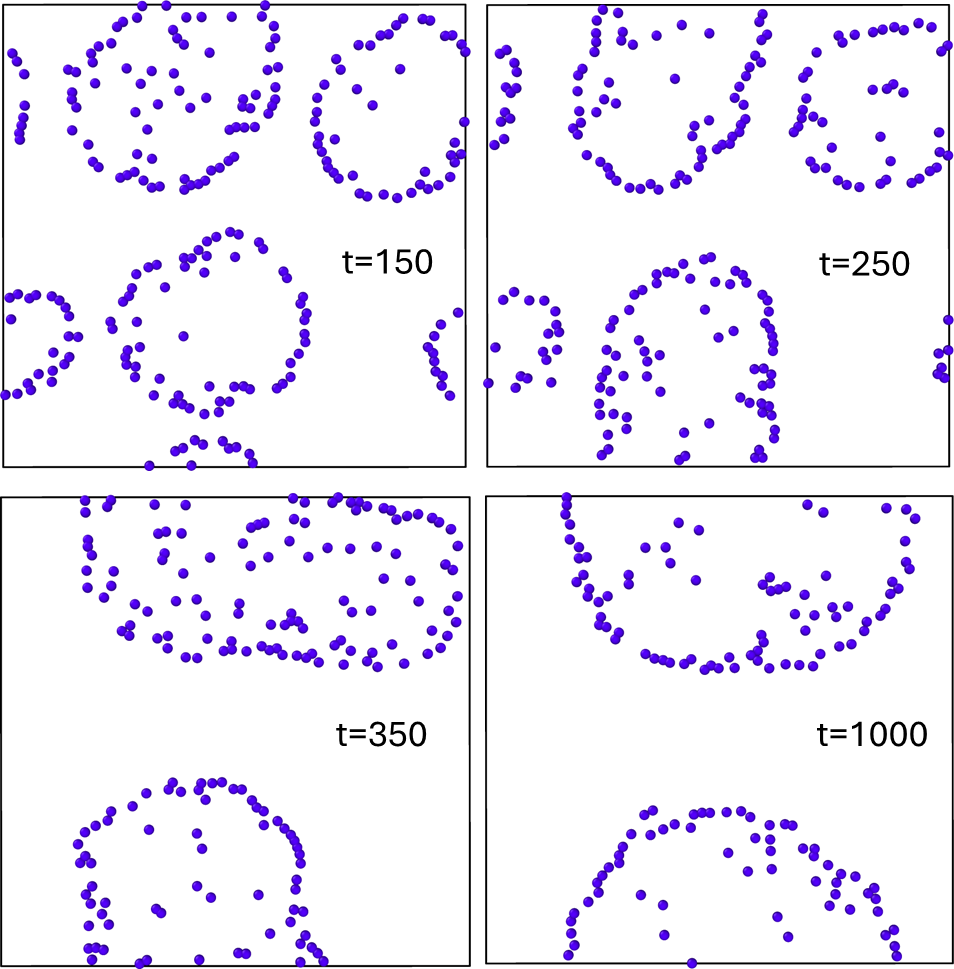}
	\caption{\label{fig:AC_circles} Merging of active circles as a function of time for the anti-correlated case at $\rho_c=0.3$ and $b=30$ (passive particles are not shown).}
\end{figure}

The location of each transition in the phase diagram can be estimated via visual inspection and conventional statistical physics measurements.
$t_1$ was estimated via visual inspection since it coincides with formation of holes with active gas particles. 
At all activities checked $t_1$ occurs near $\rho_c=0.2$. 
As activity increases, lower critical density is required for $t_1$. 
$t_2$ coincides with formation of hexatic phase detected by a split in the second peak of the radial disribution function corresponding to passive particles: $g_{pp}(r)$. 
See $g_{pp}(r)$ in Fig.S3 of SM. 
At all activities checked $t_2$ occurs near $\rho_c=0.35$. As activity increases, lower $\rho_c$ is necessary for the hexatic transition. 
Thus, activity creates stronger pressure on the passive phase to form hexatic ordering. 
$t_3$ occurs when fraction of active particles exceed 0.9, $\phi_a>0.9$. 
\\Thus, according to the criteria mentioned above, I delineated the phase diagram in Fig.\ref{fig:AC_DIAGRAM}. 
Active circles appear in the gray region of this diagram. The dynamics of the formation of circular clusters show that small circles merge into a single big circle, e.g., in Fig.\ref{fig:AC_circles} at $b=30$ and $\rho_c=0.3$. 
\section{Discussion}
I studied the emerging phases in two symmetric cases of quorum sensing and demonstrated the fundamental difference between correlated and anti-correlated density dependent motility. 
In the correlated case, high activity promotes fluid-fluid phase separation, while low activity promotes an active fluid state. 
At intermediate range of parameters, there are colonies of active swimmers without hexatic ordering. 
In the anti-correlated case, there is typically a coexistence of a compact passive phase and active gas. 
At high activity the passive solid is hexatic while at low activity it is amorphous. 
In some conditions, the active gas form circles that merge into a single circle.
\\ There are several direction for further investigation. 
One direction is simulating a model with several motilities and more than a single critical density. 
For example, in a model with two critical densities: $\rho^A_c$ and $\rho^B_c$. 
Then, as local density increases, motility changes twice: $b_0\rightarrow b_1 \rightarrow b_2$ where $b_0,b_1,b_2$ can have different values, such that motility goes: up-up, down-down, up-down, or down-up. 
Such model may produce new emerging phases. Another direction is to take DDM with a continous motility function. 
In addition, the proposed models can be studied in three dimensions. 
It would be interesting to see how dimensionality will affect the phase diagram, and more specifically if and how the active circles phase would appear in 3D. Taking non-spherical particles is another possibility that can capture shape effects that exist in biology. It was shown that rods with quorum sensing can form asters and stripes \cite{Velasco2018}.
\\ There are a few limitations in this work. Due to limited computational resources, I studied that model in a single system size. 
I predict that finite size effects are small and my system is large enough to produce the qualitative behavior. 
It was shown in other works that in two-dimensions active fluids can be hexatic with or without biological relevance \cite{HEX1,HEX2}. 
For different choice of parameters my model may produce hexatic active phase. 
\\Therefore, generic studies of quorum sensing remain relevant for the multidisciplinary academic community and can help us understand better social behaviour that is sensitive to density. 
\section{Acknowledgments}
I.A. acknowledges early discussions with Marjolein Dijkstra and Gil Ariel, further discussions with Clement Nizak, and the support of his postdoctoral advisor Erdal C. Oguz. 
This work was partially self-funded and partially funded by the International Young Scientist Fellowship of Institute of Physics, Chinese Academy of Sciences under the Grant No. 202506.
\bibliography{references}
\end{document}